\documentclass[11pt]{article}
\RequirePackage{graphicx,color,verbatim,multirow,pgf,pgfarrows,pgfnodes}

\usepackage{array}
\newcolumntype{^}{>{\currentrowstyle}}

\usepackage{longtable,tabu,amssymb}

\usepackage[default]{jasa_harvard}    
\usepackage{JASA_manu}

\usepackage{multirow}

\newcount\Comments  
\newcommand{\kibitz}[2]{\ifnum\Comments=1\textcolor{#1}{#2}\fi}

\begin{document}

\title{Incorporating Hierarchical Structure Into Dynamic Systems: An Application Of Estimating HIV Epidemics At Sub-National And Sub-Population Level}

\author{Le Bao$^1$, Ben Sheng$^1$, Xiaoyue Niu$^1$, Yuan Tang$^1$,\\ 
Tim Brown$^2$, Peter D. Ghys$^3$, Jeffrey W. Eaton$^4$.\\
Department of Statistics, Penn State University, University Park, PA, USA$^1$\\
Population and Health Studies, East-West Center, Honolulu, HI, USA$^2$\\
Strategic Information and Evaluation Department, UNAIDS, Geneva, Switzerland$^3$\\
Department of Infectious Disease Epidemiology, Imperial College London, London, UK$^4$\\
email: \texttt{lebao@psu.edu} }

\maketitle

\begin{center}
\textbf{Abstract}
\end{center}
Dynamic models have been successfully used in producing estimates of HIV epidemics at national level, due to their epidemiological nature and their ability to simultaneously estimate prevalence, incidence, and mortality rates. Recently, HIV interventions and policies have required more information at sub-national and sub-population levels to support local planning, decision making and resource allocation. Unfortunately, many areas and high-risk groups lack sufficient data for deriving stable and reliable results, and this is a critical technical barrier to more stratified estimates. One solution is to borrow information from other areas and groups within the same country. However, directly assuming hierarchical structures within the HIV dynamic models is complicated and computationally time consuming. In this paper, we propose a simple and innovative way to incorporate the hierarchical information into the dynamic systems by using auxiliary data. The proposed method efficiently uses information from multiple areas and risk groups within each country without increasing the computational burden. As a result, the new model improves predictive ability in general with especially significant improvement in areas and risk groups with sparse data.
\vspace*{.3in}


\section{Introduction}
\label{sect-Introduction}

Mounting an effective response to HIV/AIDS requires reliable estimates of the global scope and recent trends in the epidemic. National governments, with support from UNAIDS, use a dynamic mathematical modeling software called Spectrum to annually generate national HIV epidemic estimates \cite{Case2014}. Spectrum was applied in 163 countries in 2015. Recent estimates suggest that great progress has been made, including a 38\% decline in new infections globally since 2001, a 58\% drop in new infections in children since 2002, and a 35\% fall in AIDS-related death since 2005 \cite{UNAIDS2014}.


However, it is perceived that among these successes, certain key populations at higher risk of acquiring HIV and sub-national geographic regions with greatly elevated burden have not benefited equally. Young women are disproportionately affected by HIV in Sub-Saharan Africa, accounting for 71\% of all infections among adolescents. Prevalence reviews have shown that sex workers and men who have sex with men have ten-fold or higher prevalence compared to the population as a whole \cite{Beyrer2012,Baral2012}. People who inject drugs account for an estimated 30\% of new HIV infections outside of Sub-Saharan Africa. And within high burden countries, geographic variations by province and district can be large, e.g., in Zambia HIV regional HIV prevalence varied by a factor of three from 6.4\% to 18.2\% in 2013 \cite{ZambiaDHS2015}.”

The existing tools for generating estimates involve fitting a dynamic epidemic model called the Estimation and Projection Package (EPP) to national surveillance data about HIV prevalence using a Bayesian framework \cite{Alkema2007}. The dynamic system modeling framework has several advantages including its epidemiological nature, ability to make short-term projection, and simultaneous estimation of prevalence, incidence, and mortality rates. It is well known that HIV epidemics can spread heterogeneously across sub-national areas and risk groups. Nevertheless, current approaches to monitor the epidemic do not always adequately capture this heterogeneity. Currently, the sub-national estimations are done by applying the epidemic models \textit{independently} using only data from within the area/group \cite{Mahy2014}. However, the availability and quality of the data vary widely. For areas/groups with sparse data, the models produce inaccurate results with large uncertainty bounds \cite{Lyerla2008,Calleja2010}. 

One solution is to assume that the parameters of the epidemic models are correlated among areas, groups, or even countries in a hierarchical framework, thereby efficiently borrowing information from other areas and groups with similar epidemics. However, estimating parameters in the HIV epidemic models is already time consuming because of the dynamic systems that are used to construct the historic trends of multiple quantities of interest, such as the number of people living with HIV, the number of new infections, and the number of HIV deaths. Directly assuming hierarchical structures and modeling the joint distribution of parameters in multiple areas and risk groups further challenge the estimation and computation. 

In this paper, we propose a simple and innovative way to incorporate the hierarchical information into the dynamic systems by using auxiliary data. The proposed method efficiently uses information from multiple areas and risk groups within each country without increasing the computational burden. In Section 2, we describe the dynamic models used to produce national estimates. In Section 3, we introduce our method that extends the sub-national / sub-population estimation to include the hierarchical structure for sharing information across areas and risk groups. In Section 4, we evaluate the model performance via test data prediction, and present results for Nigeria and Thailand. In Section 5, we offer conclusions and discussion for future work.

\section{The HIV estimation model}
\label{sect-EPP}
The most commonly used model to fit and project HIV epidemics is the Estimation and Projection Package (EPP), which is the primary source of incidence estimates in Spectrum, the UNAIDS provided software used by most countries. EPP is based on a susceptible-infected (SI) epidemiological model. The dynamic model is as follows:
\begin{equation}
\left\{\begin{array}{ccc}
\frac{dZ(t)}{dt} & = & E(t) - r(t) \rho(t) Z(t) - \mu(t) Z(t) - a_{50}(t)Z(t) +  M(t)Z(t), \\
\frac{dY(t)}{dt} & = & r(t) \rho(t) Z(t) - \textup{HIVdeath}(t) - a_{50}(t)Y(t) +  M(t)Y(t). \\
\end{array}\right.
\label{eqn:ode}
\end{equation}
The model represents the adult (aged 15-49 year) population stratified into the susceptible population $Z(t)$ and infected population $Y(t)$ at time $t$, such that the total adult population size is $N(t) = Z(t) + Y(t)$. The transmission rate over time from infected to uninfected adults is modeled as a flexible smooth function $r(t)$. The functional form for $r(t)$ is either modeled by B-splines in the `r-spline' model \cite{Hogan2010} or a parametric autoregressive model in the r-trend model \cite{Bao2010rflex,Bao2012rtrend}. $E(t)$ is the number of new adults entering the population at age 15, $\rho(t)=Y(t) / N(t)$ is the prevalence, $\mu(t)$ is the non-HIV death rate, and $\textup{HIVdeath}(t)$ is the number of deaths among infected individuals. $a_{50}(t)$ is the rate at which adults exit the model after attaining age 50, and $M(t)$ is the rate of net migration into the population. For some high-risk groups, such as people who inject drugs and sex workers, entrance and exit into the population is also determined by the average duration of staying within the high-risk population. Given a small initial seed prevalence at the start of the epidemic, the model simulates a temporal trend of prevalence, incidence, and HIV mortality rates.

The main data source for EPP model estimation in the general population is sentinel surveillance of HIV prevalence among pregnant women attending antenatal clinics (ANC). These consist of the number of infected women $Y_{it}$ and the number of tested women $N_{it}$, for a given year $t$ and a given clinic $i$. Where available, model estimation also uses population prevalence estimates for the entire population from nationally representative household surveys. 

To account for multiple sentinel sites in the same region, the observed clinic-level prevalence and the population prevalence -- $\rho(t)$ -- derived from the dynamic system (\ref{eqn:ode}) are linked through a mixed effects model. In addition, the prevalence is transformed to be put into a linear mixed effect model so that the random effects and their variance can be integrated out analytically under the normality assumption, and thus simplify the parameter estimation \cite{Alkema2007,Bao2012rtrend}. 
\begin{eqnarray}
W_{it}&=&\Phi^{-1}(\rho_t)+\alpha+b_i+\epsilon_{it}, \nonumber\\
b_i&\sim& N(0,\sigma^2), \label{eqn:ranef} \\
\epsilon_{it}&\sim&N(0,\nu_{it}) \nonumber
\end{eqnarray}
where $\Phi^{-1}(.)$ is the inverse cumulative distribution function of the standard normal distribution, $W_{it}=\Phi^{-1}(\frac{Y_{it}+0.5}{N_{it}+1})$, $\alpha$ is the bias of ANC data with respect to prevalence data from national population-based household surveys (NPBS), $b_i$ is the site-specific random effect, $\sigma^2$ is assumed to have an inverse-Gamma prior which gets integrated out in the likelihood evaluation, and $\nu_{it}$ is a fixed quantity that depends on the clinic data and approximates the binomial variation.

Finally, a Bayesian framework is used to bring in the prior information of the key parameters such as the starting time of the epidemic. Model parameters are estimated with posterior approximation via Incremental Mixture Importance Sampling (IMIS) \cite{Raftery2010}. The final estimates reflect three sources of information: the prevalence data, the prior knowledge, and the epidemic trends inferred by the SI model in Equation \ref{eqn:ode}. As a result, the EPP model has several advantages for estimating the HIV epidemics. Its basis in an dynamic epidemic model lends theoretical credibility to the resulting epidemic inference, and the model intrinsically represents the relationship between key epidemiological processes, enabling internally consistent estimates about multiple quantities of interest including prevalence, mortality, and incidence, about which there is no directly observable data. Finally, the dynamic model is linked to a statistical model so that it can produce uncertainty estimates and projection intervals for all quantities of interest. We will continue to use the EPP model as a building block to produce sub-national and sub-population epidemic estimates.

\section{Incorporating hierarchical structure into estimating sub-epidemics}
\label{sect-Hierarchical models}

The current approach to extend the EPP model to fit sub-epidemics is to run the EPP model for each sub-national area and risk group combination independently using only data from within the sub-epidemic.  However, many sub-epidemics lack sufficient data for deriving stable and reliable results (see Figure \ref{fig:Nigeria} (d) as an example where the uncertainty is too large for the early period of the epidemic without sufficient data). Our goal is to improve the accuracy of the results in areas and groups with sparse data, while retaining the simplicity of the epidemic model and not increasing the computational burden. In previous work, we have demonstrated the application of a full hierarchical model for joint parameter estimation across regions \cite{Bao2015a}. However, this is burdensome for a computationally expensive dynamic model and likelihood, and may prove to be infeasible for some routine applications.  The main idea proposed here is very simple: we propose to fit a generalized linear mixed model (GLMM) to data from all sub-national areas and risk groups, and incorporate the fitted prevalence into the EPP model estimation through auxiliary data to efficiently approximate the information that would be added through a hierarchical structure. The approach can be outlined as follows:
\begin{enumerate}
\item For a particular country, we pool the data from all areas and risk groups together and fit a GLMM with a non-parametric flexible time trend and random effects for areas, sites, and risk groups to represent the hierarchical structure.
\item We then use the predictive distribution of the sub-epidemic prevalence from the GLMM to create auxiliary data. The precision of the auxiliary data is determined by the uncertainty estimate from GLMM. 
\item Finally, we add the auxiliary data to the original data and fit the EPP model separately for each sub-epidemic as before. The resulting prevalence, incidence, and mortality estimates can all be derived the same way as before, while they now contain the information from other areas that would be included through the hierarchical structure.
\end{enumerate}
The details of the model are described in the following subsections.  


\subsection{Generalized linear mixed models (GLMM)}

In the first step, we model the prevalence data without using the epidemic model, so that the information can be efficiently pooled across areas. GLMM is a natural choice to model such data which takes the hierarchy or even spatial dependence into consideration. Instead of using a dynamic model, we model prevalence trends as functions of time over the period in which data were observed. Spline models introduce great flexibility with fixed degree, and relieve the pressure of over-fitting \cite{Wold1974,Wahba1990}. With some experiments, we choose the natural cubic spline model with equally spaced knots to describe the time trends. The natural spline uses boundary constraints to stablize the estimates near the data boundary, which is helpful when few observations are available in the begining or end year of the data.


We first apply GLMM to countries in which the overall prevalence is high and the epidemic is not confined to particular subgroups. Therefore, there is only one risk group to be considered, that is, the whole adult population of the country. The proportions of HIV+ cases among antenatal clinic (ANC) attendees observed over years are often used as a proxy for the time trend of adult prevalence in general population. In a typical epidemic with multiple areas, and multiple surveillance sites within each area, let $a$, $i$, $t$ indicate area, site, and time respectively. It is expected that prevalence data collected from the same site or different sites within the same area are correlated. We test GLMM specifications of varying complexity, and the following model is recommended:
\begin{equation}
\left\{\begin{array}{ccc}
Y_{ia}(t) & \sim & \mbox{Binomial}(n_{iat},\rho_{ia}(t)) \\
\mbox{logit}( \rho_{ia}(t) )& = &\beta_{0}+\beta_1 f_1 (t) + \beta_2 f_2 (t) + \ldots +
\beta_{K} f_K (t) + \\
&& b_{0a} + b_{1a} f_1 (t) + b_{2a} f_2 (t) + \ldots + b_{Ka} f_K (t) + b_{i(a)},
\end{array}\right.
\label{eqn:GLMM1}
\end{equation}
where $f_{1}, f_{2}, \cdots, f_{K}$ are the basis functions of the natural cubic spline; $\beta$\rq{}s are fixed effects that determine the time trend at the national level; $b$\rq{}s are the random effects at area and site levels. It reflects the hypothesis of different but similar area-level epidemic trends by including area-level random intercepts and spline coefficients. 

In countries with low-level and concentrated epidemics, HIV has spread rapidly in the key populations that are most likely to acquire and transmit HIV, but is not well established in the general population. There is no set of representative data that can be used for the general population, and the estimation is done within each individual group. The sub-epidemics could be quite different across high-risk groups such as people who inject drugs, sex workers, clients of sex workers, and men who have sex with men. Therefore, we suggest introducing sub-population-specific intercepts and spline coefficients in Equation \ref{eqn:GLMM1}, and treating them as fixed effects. Other components remain the same, so that the area-specific and site-specific parameters are shared by all sub-populations, e.g. the area with relatively high HIV prevalence among people who inject drugs also has relatively high HIV prevalence among sex workers, compared other areas. Let $g$ be the index of risk groups. The following hierarchical model is suggested:
\begin{equation}
\left\{\begin{array}{ccc}
Y_{iag}(t) & \sim & \mbox{Binomial}(n_{iagt},\rho_{iag}(t)) \\
\mbox{logit}( \rho_{iag}(t) )& = &\beta_{0g}+\beta_{1g} f_1 (t) + \beta_{2g} f_2 (t) + \ldots +
\beta_{Kg} f_K (t) + \\
&& b_{0a} + b_{1a} f_1 (t) + b_{2a} f_2 (t) + \ldots + b_{Ka} f_K (t) + b_{i(a)},
\end{array}\right.
\label{eqn:GLMM2}
\end{equation}
where $\beta$\rq{}s are group-specific fixed effects, $b$\rq{}s are the random effects at area and site levels. 

For both models, we assign diffuse Gaussian priors to the fixed effects, and use the Monte Carlo Markov Chain (MCMC) algorithm implemented in the R package -- {\it MCMCglmm} \cite{Hadfield2010}, to approximate the posterior distribution of area-specific prevalence, $\rho_a(t)$, with the site effects being marginalized. We run 105,000 iterations with 5,000 burn-in and 20 thinning interval, and collect a final set of 5,000 posterior samples. The convergence of Markov chains is checked based on trace plots and auto-correlation function (ACF) plots \cite{Brockwell2013}.

To verify our model choices, we compare the above models with the alternative hierarchical structures and prior distributions. The model performance is evaluated by using (1) the deviance information criterion (DIC) \cite{Spiegelhalter2002,Spiegelhalter2014}; (2) the mean absolute error (MAE) of the test dataset prediction. This ensures that our proposed models either perform the best or nearly the best among all candidate models. Detailed model specifications and results are provided in the Appendix A.





\subsection{Incorporating hierarchical information into EPP model estimation}
GLMM utilizes data information efficiently by assuming similarity of the time trends (spline coefficients) while allowing heterogeneity across areas and risk groups, so that the prevalence trend in a specific area/risk group will be determined not only by its own data but also data from other areas and groups. However, it ignores the epidemic model and only provides inference for HIV prevalence trends within the period in which data were observed. Ideally, one could have replaced the spline function with the prevalence trend produced by the epidemic model in Equation \ref{eqn:ode}. The difficulty is that estimating parameters in dynamic models is time consuming due to the lack of analytic solutions. Estimating multiple dynamic systems -- one for each area -- will further increase the computing cost and makes it unfeasible for a country with many areas.

Instead, we use the posterior distribution of the prevalence from the GLMM to create auxiliary data to add onto the original data for each area and each risk group. The auxiliary data can be viewed as an approximation for the prior distribution of the prevalence, and contains the hierarchical information from the GLMM. More specifically, for area $a$ and risk group $g$ in year $t$, we denote the posterior estimate of $\rho_{ag}(t)$ as $\tilde{\rho}_{ag}(t)$ with mean $\mu_{agt}$ and variance $v_{agt}$. We approximate this posterior information by converting each to a binomial observation with prevalence $\mu_{agt}$ and sample size $n_{agt}=\frac{\mu_{at}(1-\mu_{at})}{v_{at}}$. This binomial formulation facilitates straightforward inclusion of auxiliary information into the existing EPP likelihood function and intuitive adjustment of the \lq{}strength\rq{} of the auxiliary prevalence information by scaling the auxiliary data sample size. The total auxiliary data sample size for the sub-epidemic of area $a$ and risk group $g$ can be set to any predetermined number $K$ by scaling $n_{agt}$ proportionally, so that the relative strength of $n_{agt}$ remains the same within each sub-epidemic. $K=0$ corresponds to fitting Spectrum/EPP to data within area $a$ and group $g$ without borrowing information from other areas or groups. As $K$ increases, the prior prevalence distribution becomes more informative. In two illustrative examples below, we determine the optimal default value of $K$ using cross-validation.

\subsection{Model validation}
We evaluate the model performance by its out-of-sample prediction as outlined below:
\vspace{-0.05in}
\begin{enumerate}
\item For each sub-epidemic, randomly partition observations into training and test sets. 
\vspace{-0.08in}
\item Apply GLMM to the training data of all sub-epidemics.
\vspace{-0.08in}
\item Generate auxiliary data from the predictive distributions of GLMM.
\vspace{-0.08in}
\item For each sub-epidemic, apply the EPP model to the training data with the addition of auxiliary data of size $K=0 \textup{ (original EPP)}, 50, 100, 200, 500, 1,000, 2,000, 5,000, 10,000$.
\vspace{-0.08in}
\item For each sub-epidemic, compare the observations in the test data with the corresponding predictions from the model. Let $w_{it}$ be the observed prevalence of site $i$ and year $t$ on the probit scale in the test data, and $w_{it}^{(j)}$ be the $j$th posterior predictive sample from the model. The mean absolute error (MAE) is defined as $|w_{it}-w_{it}^{(j)}|$ averaged over all the posterior samples, sites, and years in the test data. This single-number criterion takes both the bias and the uncertainty of the predictive distribution into consideration. In addition, we evaluate the uncertainty estimates by the coverage and the average width of the 95\% prediction intervals.
\end{enumerate}
 
\section{Results}
\label{sect-Results}
As illustrative examples, we present results for two countries---Nigeria and Thailand. Nigeria represents countries with high (\textgreater 1\%) HIV prevalence, where pregnant women prevalence (as measured in antenatal clinics) has been used as the indicator of the epidemic trend, together with available data from national population-based surveys. There are about 49 such countries, most of them in sub-Saharan Africa. Thailand represents epidemics where the HIV prevalence is lower and largely concentrated among specific sub-populations, including sex workers and their clients, gay and other men who have sex with men, and people who inject drugs. For each example, we compare the proposed model with the existing approach, i.e. independent fitting of EPP. 

\subsection{Nigeria Example}


Nigeria consists of 36 states and the Federal Capital Territory Abuja with surveillance data starting from 1992. The data quality varies across states: the number of years for which ANC data is available ranges from 6 to 9; the number of clinic sites per state ranges from 2 to 8. The number of tested individuals of each area ranges from 3,482 to 10,789 with a median of 4,850.

Following the procedure described in Section 3.3, we partitioned the data using a training-test ratio of 1:1 at each site. The auxiliary data was generated by applying the GLMM specified in Equation \ref{eqn:GLMM1} to the training datasets of 37 areas. For the test dataset of each area, we calculated the mean absolute error (MAE) with varying auxiliary data sample size, $K=50, 100, 200, 500, 1,000, 2,000, 5,000$, $10,000$, as well as the MAE of the original EPP estimate without auxiliary information ($K=0$). The random training-test splitting and the corresponding MAE evaluations ware repeated 10 times, and thus, there were 370 test datasets from the 10 training-test splits and 37 areas. 

\begin{figure}[!h]
\begin{tabular}{cc}
\begin{minipage}{8cm}
\includegraphics[width=8cm]{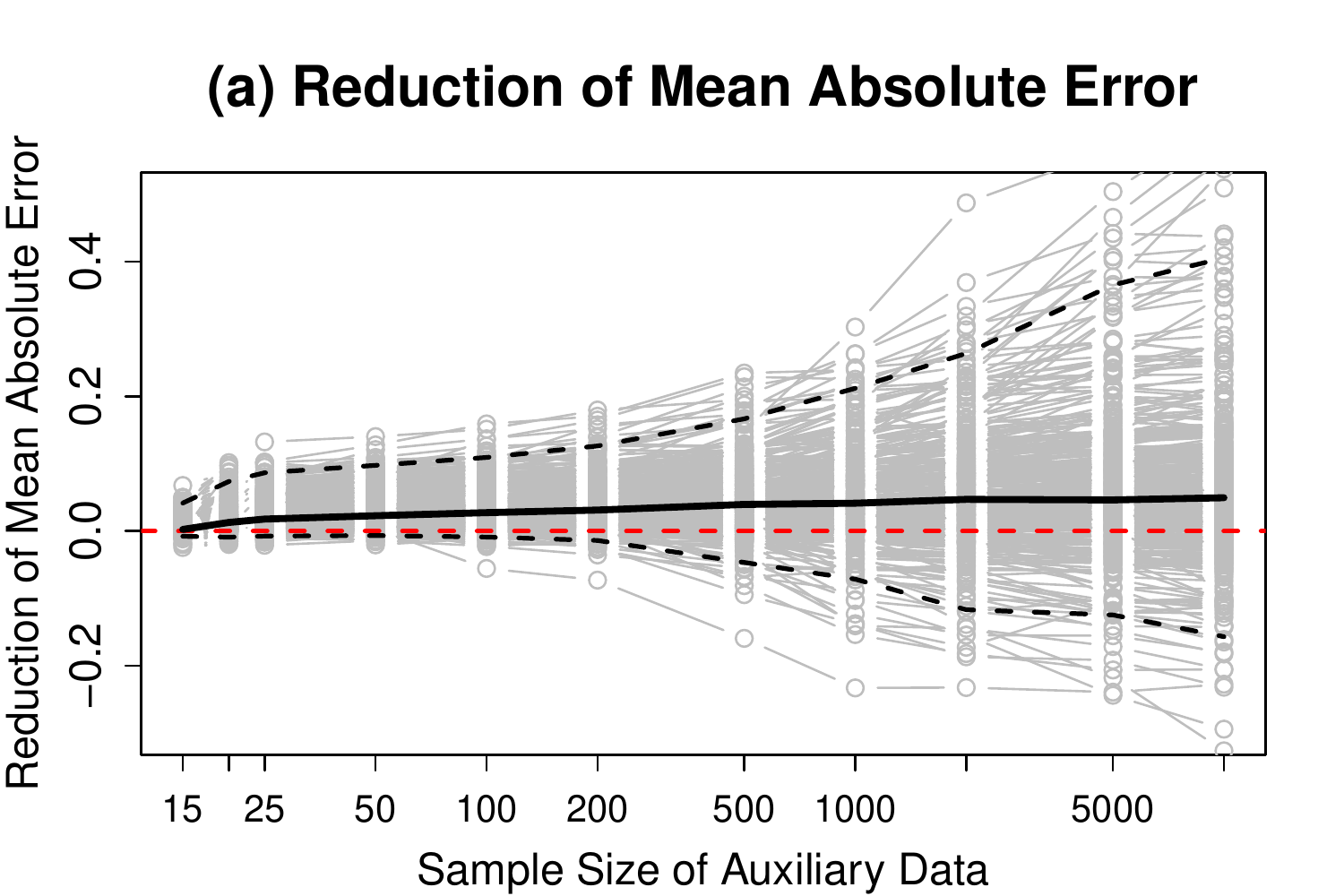}
\end{minipage}
&
\begin{minipage}{8cm}
\includegraphics[width=8cm]{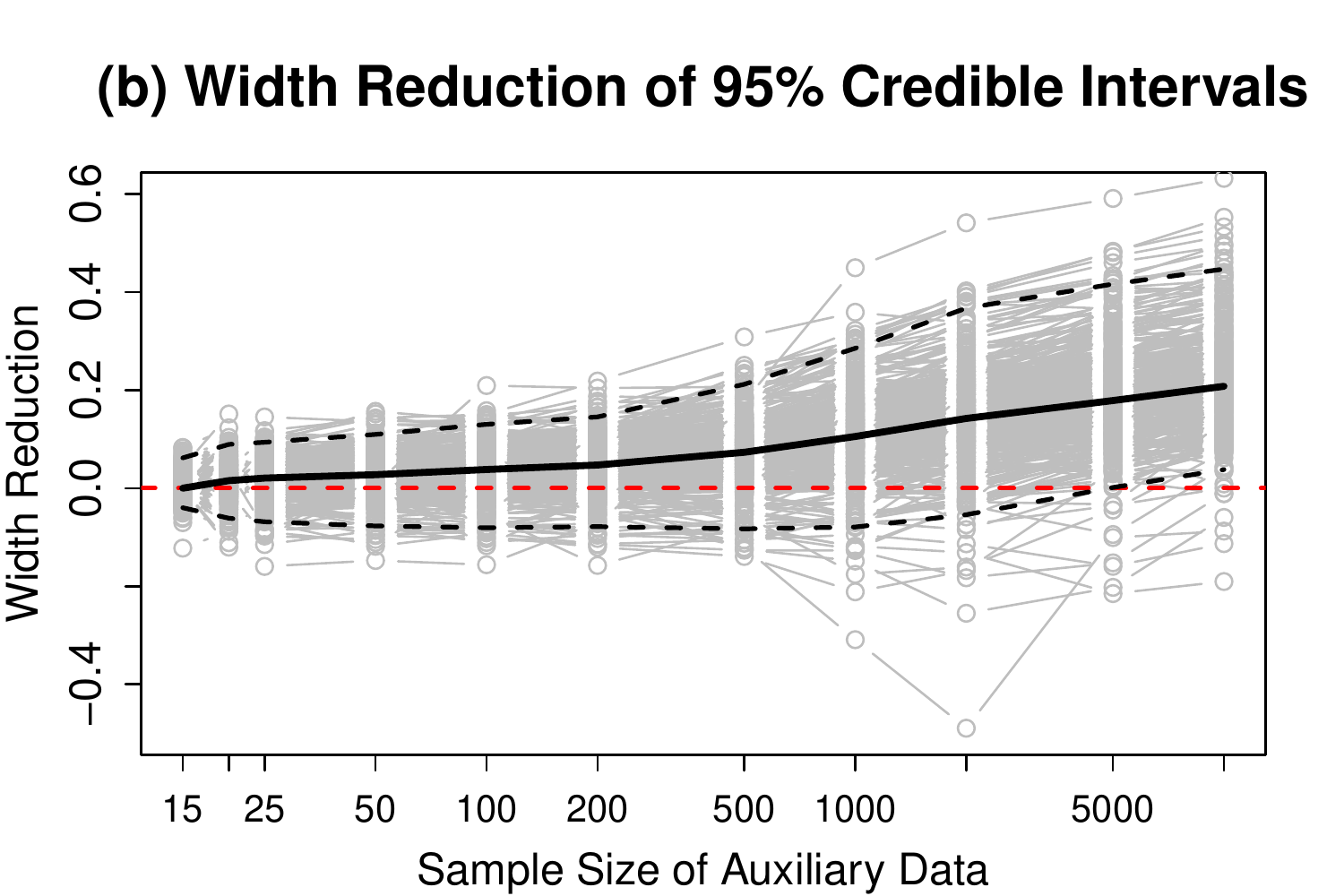}
\end{minipage}
\\
\begin{minipage}{8cm}
\includegraphics[width=8cm]{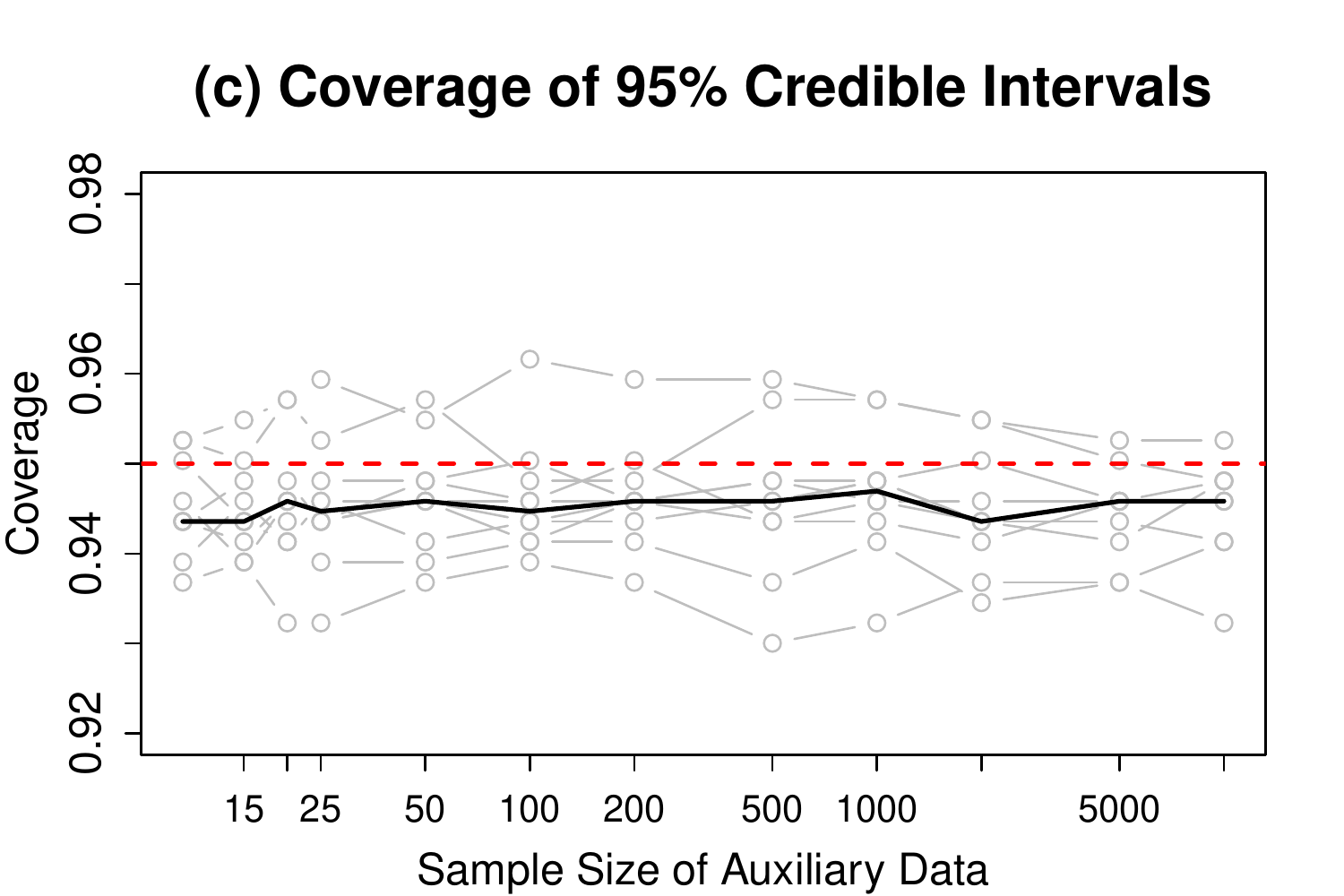}
\end{minipage}
&
\begin{minipage}{8cm}
\includegraphics[width=8cm]{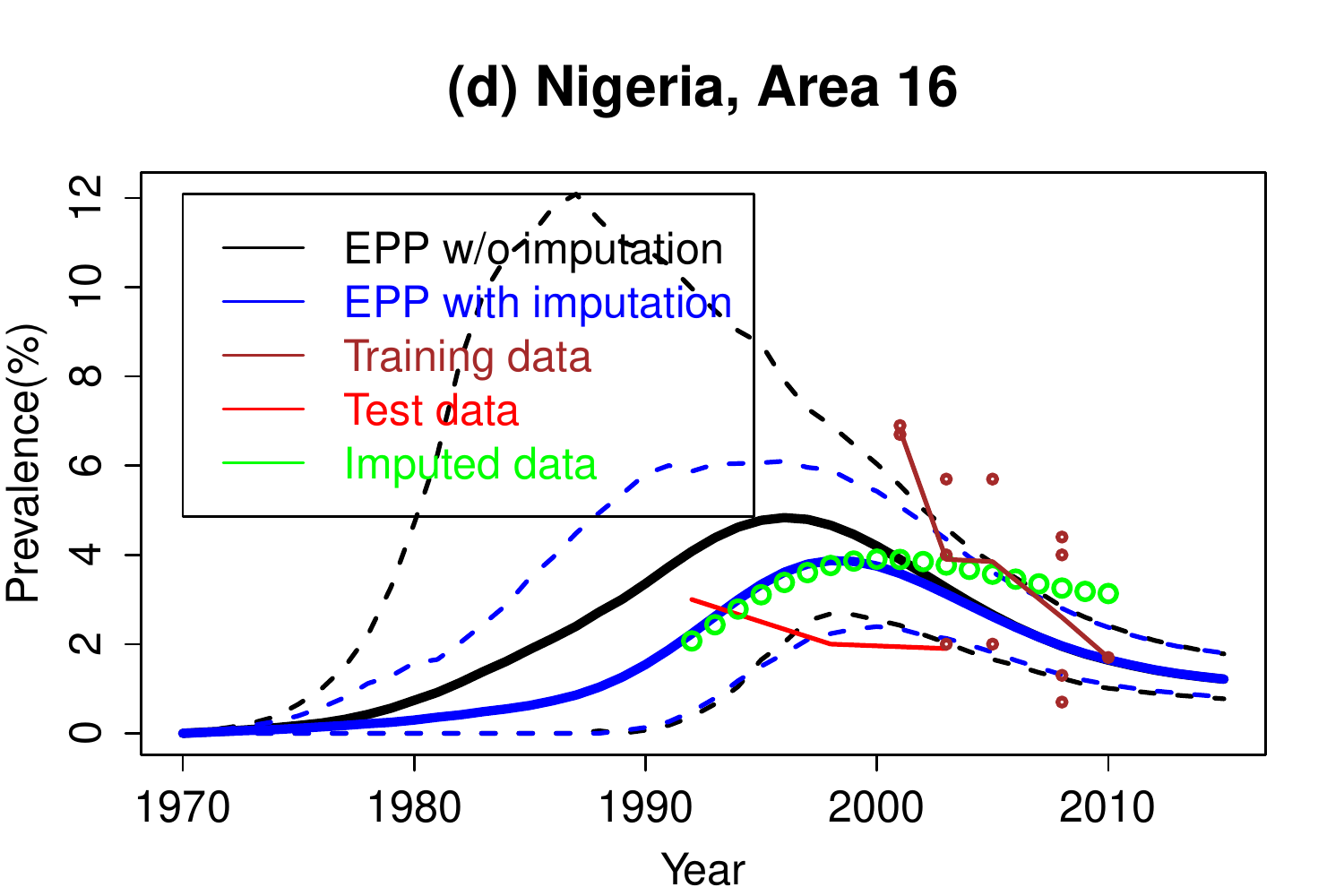}
\end{minipage}
\\
\end{tabular}
\caption{\footnotesize{Nigeria example. In (a) and (b), the gray curves show the trends of the relative reduction of each quantity with respect to the original EPP model ($K=0$) as $K$ increases up to 10,000 for each area; the black solid curve shows the median among 370 test datasets (37 areas and 10 training-test splits); the black dashed curves indicate the 2.5th and 97.5th quantile; the red dashed line indicates 0 reduction. In (c), the gray curves show the coverage of 95\% credible interval for each training-test split; the black solid curve shows the median among 10 training-test splits; the red dashed line corresponds to 95\%. In (d), the black curves show the posterior median and 95\% credible interval of prevalence trends estimated from EPP without using auxiliary data; the blue curves show the posterior median and 95\% credible interval of prevalence trends estimated from EPP by using auxiliary data of sample size $K=200$; the brown curve shows the average prevalence in training data; the red curve shows the average prevalence in test data; the green dots show the auxiliary data.}}
\label{fig:Nigeria}
\end{figure}

For each test dataset, we calculated the MAE reduction of the proposed model relative to the original EPP as $\frac{\mbox{MAE}_0-\mbox{MAE}_K}{\mbox{MAE}_0}$. Figure \ref{fig:Nigeria} (a) illustrates the relative MAE reduction with varying values of $K$, and each curve represents one test dataset. The black solid curve shows the median among the 370 test datasets; and the black dashed curves correspond to 2.5th qunatile and 97.5th quantile. As $K$ increases from 15 to 200, the median MAE reduction increases from 0.2\% to 3.1\%, and MAE reductions are positive for most test datasets. For $K$ between 500 and 10,000, the median reduction further increases up to 5.0\%. However, the variation of MAE reductions across test datasets becomes larger, and there are a few test datasets whose MAE reduction are less than -20\%, or, equivalently, whose MAE increase by 20\% or more after including auxiliary data. To prevent a large increase of MAE in any test dataset, we recommend $K=200$ at which the MAE reductions range from -7.2\% to 18.0\% with median 3.1\%. Among 370 test datasets, 337 have positive reductions, and only 13 have negative reduction of more than 1\%.


Similarly, we examine the relative reduction of the average width of the 95\% credible intervals for HIV prevalence estimates. Figure \ref{fig:Nigeria} (b) shows that including the auxiliary data narrows the credible interval with positive width reductions. In addition, Figure \ref{fig:Nigeria} (c) shows the coverage of 95\% credible intervals with $K$ varying from 0 to 10,000. There are too few data points in each area to derive reliable area-specific coverage. For each training-test split, we combine the test datasets of 37 areas, and define the coverage as the proportion of all test data points that fall within the 95\% credible intervals. There are 10 gray curves, one for each training-test split, and the black curve indicates their median. Those proportions are fairly close to 95\% and have no trend over $K$. Therefore, we conclude that introducing the auxiliary data improves the test data prediction accuracy and reduces the range of uncertainty bounds without diminishing their coverage. 



Finally, Figure \ref{fig:Nigeria} (d) compares the estimated area-level prevalence trends from different models in one training-test split as an illustrative example. The training data (brown color) are available between 2001 and 2010. The black curves show the posterior median and 95\% credible interval (black dotted curves) of prevalence trends estimated from EPP without using auxiliary data, and there is a huge amount of uncertainties before 2001 due to lack of data. The auxiliary data (green color) incorporate prevalence trends from other areas in Nigeria, and suggest that the prevalence peaks in late 1990 and slowly declines since then. The blue curves show the posterior median and 95\% credible interval of prevalence trends estimated from EPP with auxiliary data of sample size $K=200$. Introducing the auxiliary data does not lead to major change in the prevalence estimates after 2000, but greatly reduces the uncertainty before 2000. The blue curves (with auxiliary data) suggest less dramatic changes of prevalence before 2000, and better agree with the test data average (red lines) than the black curves (without auxiliary data). 

\subsection{Thailand Example}
The Thai epidemic grew rapidly in the late 1980s when prevalence among people who inject drugs, female sex workers and their clients rose rapidly nationwide, shortly followed by growing prevalence among pregnant women. The epidemic is characterized by strong regional prevalence variations with the highest rates in the North and the lowest in the Northeast of the country. Thailand has an extensive HIV surveillance data set for key populations dating from 1989 when surveillance was first instituted in 14 provincial capitals. By 1991 all provinces were collecting prevalence data in multiple populations \cite{Weniger1991,Brown1994}. Due to aggressive national prevention efforts, prevalence in sex workers, clients and pregnant women began a rapid decline in the mid- to late-1990s. However, prevalence among men who have sex with men and people who inject drugs remain high.

The current epidemic estimation for Thailand is stratified into multiple sub-populations and four geographic areas: Central (26 surveillance sites), North (17 surveillance sites), Northeast (19 surveillance sites), and South (14 surveillance sites). Here, we focus on three sub-populations with relatively reliable data: pregnant women, indirect sex workers, and direct sex workers. Prevalence data are observed from 1989 to 2011. The data availability varies across regions and sub-populations; among 23 data years and 76 surveillance sites, the missing data proportions are 6.9\% for pregnant women, 36.0\% for indirect sex workers, and 43.5\% for direct sex workers. For ANC data, the average number of tested individuals is 282,699 per area. For indirect sex worker data, the average number of tested individuals is 56,796 per area. For direct sex worker data, the average number of tested individuals is 44,160 per area. 

GLMM described in Equation \ref{eqn:GLMM2} incorporates prevalence data from multiple risk groups. Since epidemic patterns are often quite different among sub-populations with various risk behaviors, we treat group-specific intercepts and group-specific spline coefficients as fixed effects. For the spatial dimension, we recommended the same hierarchical structure as in Nigeria example: an intercept and a set of national-level spline coefficients as fixed effects; site-specific intercepts, area-specific intercepts, and area-specific spline coefficients as random effects.  The proposed model is further compared with a set of alternative models in Appendix A.2.

Note that Thailand is among the countries with the highest quality of surveillance data for key sub-populations that have high-risk behaviors. In a typical country of concentrated HIV/AIDS epidemic, surveillance data for key sub-populations are often sparse due to the stigmatized nature of those sub-populations. We would like to mimic the sparse data situation when constructing the training set. For each sub-population in each area, we randomly select 3 sites and then take 3 years of data from each site as the training data, and use the remaining years of those sites as the test data. Following the procedure described in Section 3.3, and similar to the Nigeria example, we obtain the mean absolute error (MAE) for each sub-population and each area, with various auxiliary data sample sizes. Finally, we repeat the random training-test splitting and corresponding evaluations 10 times. 

\begin{figure}[!h]
\begin{tabular}{cc}
\begin{minipage}{8cm}
\includegraphics[width=8cm]{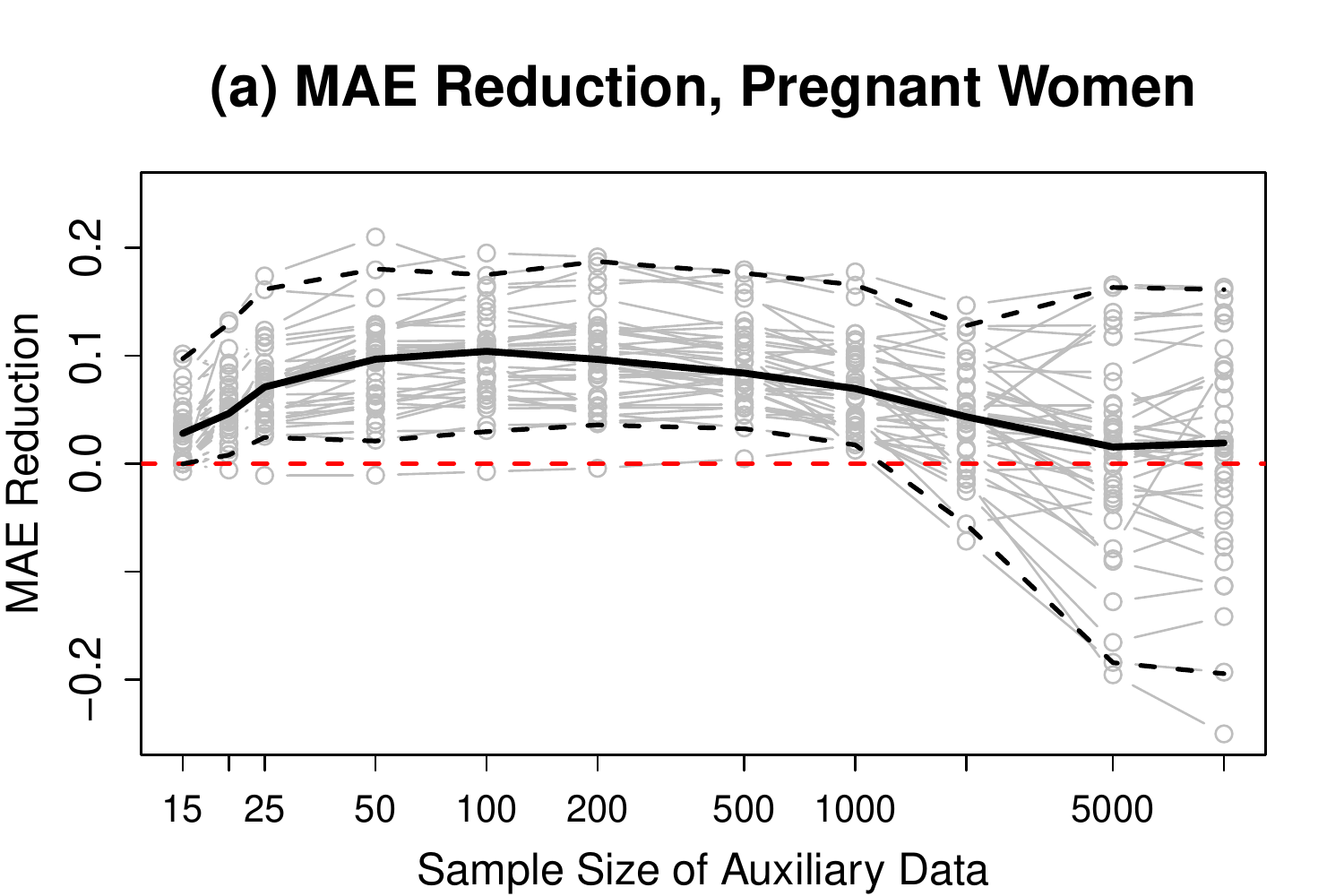}
\end{minipage}
&
\begin{minipage}{8cm}
\includegraphics[width=8cm]{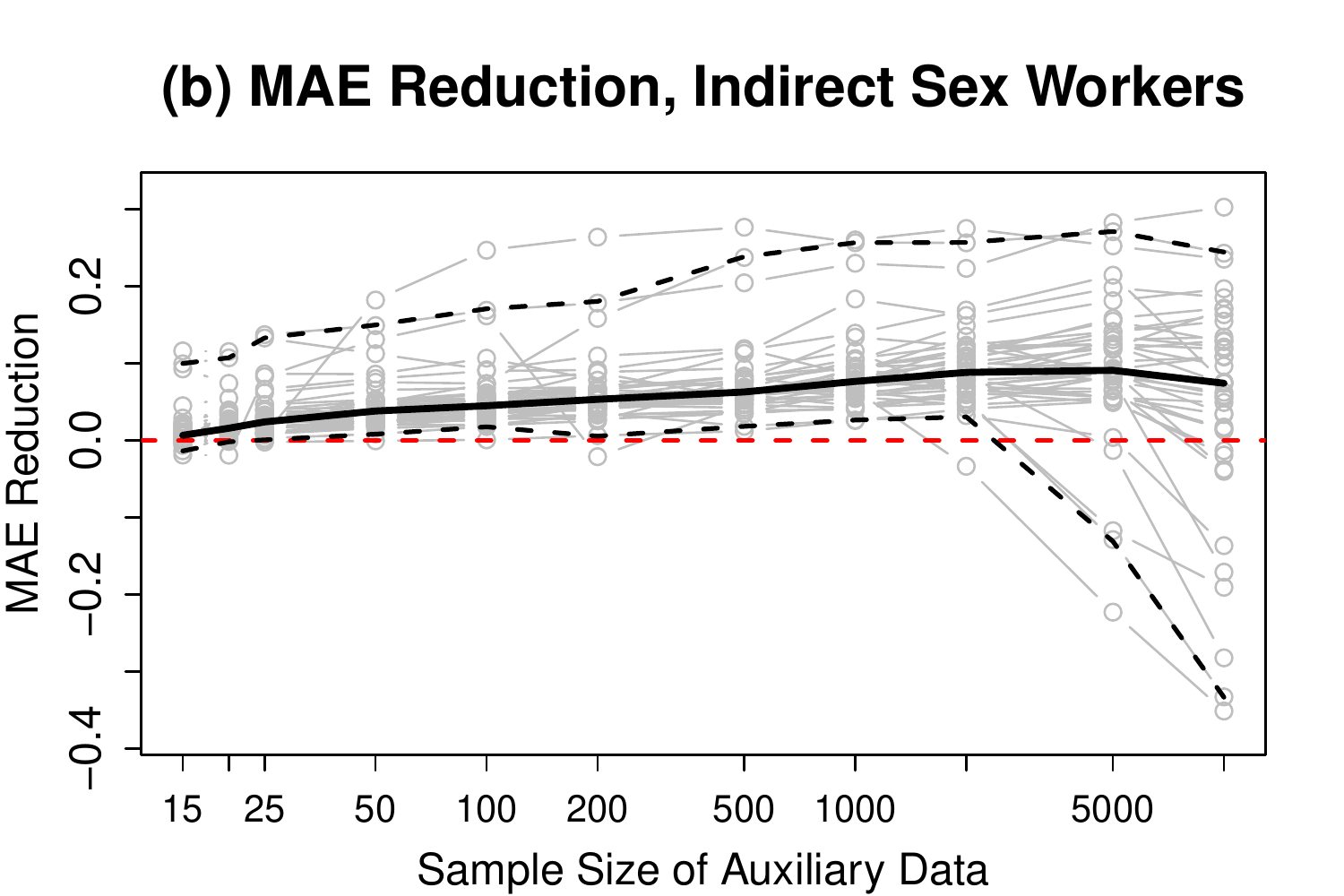}
\end{minipage}
\\
\begin{minipage}{8cm}
\includegraphics[width=8cm]{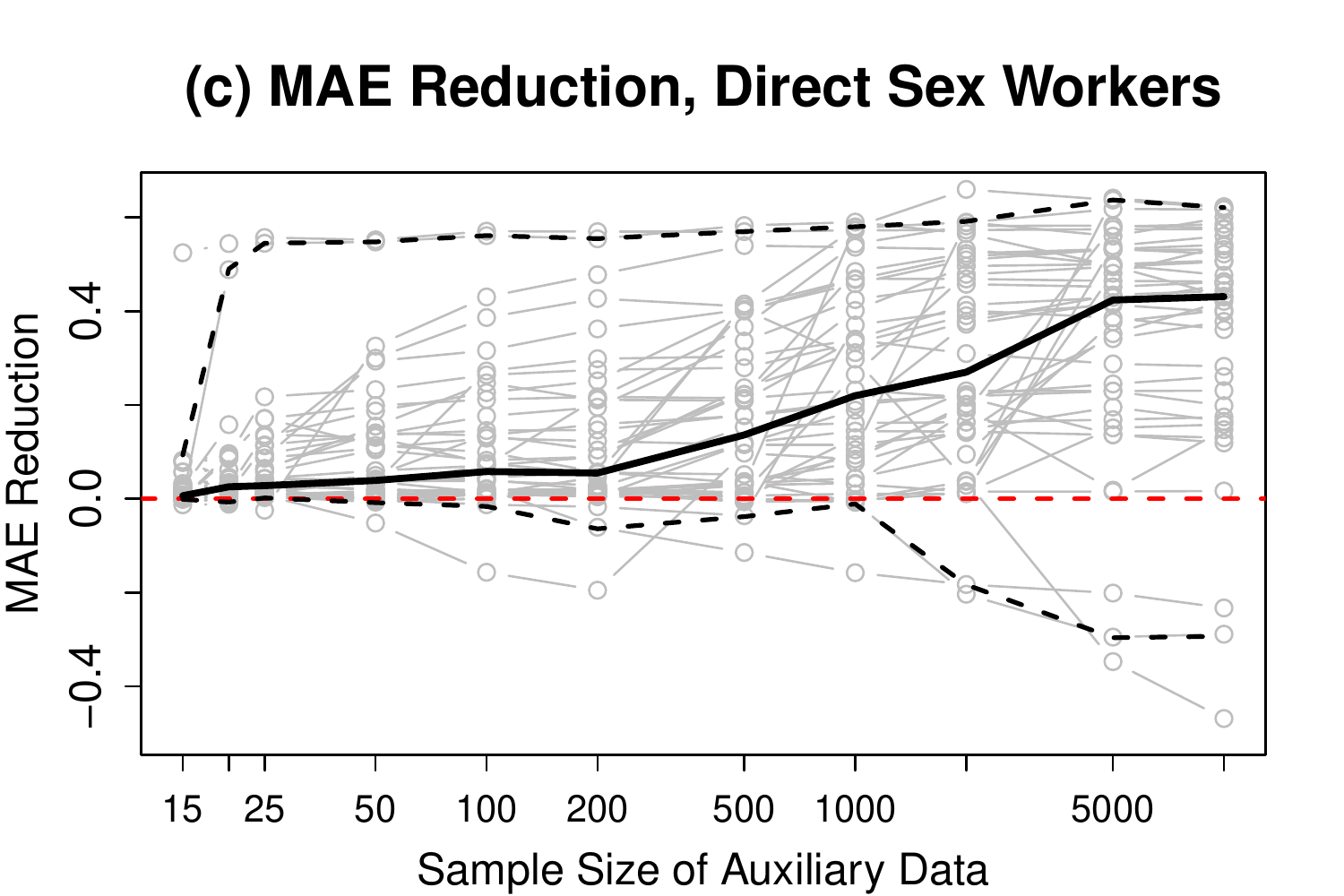}
\end{minipage}
&
\\
\end{tabular}
\caption{\footnotesize{Reduction of mean absolute error (MAE) for three sub-populations in Thailand: (a) pregnant women, (b) indirect sex workers, (c) direct sex workers. The gray curves show the trends of the relative MAE reduction with respect to the original EPP model ($K=0$) as $K$ increases up to 10,000 for each area and each training-test split; the black solid curve shows the median; the black dashed curves show the 2.5th and 97.5th quantile; the red dashed line indicates 0 reduction.}}
\label{fig:MAE_Thai}
\end{figure}

Figure \ref{fig:MAE_Thai} summarizes the reduction of MAE for three sub-populations in Thailand: (a) pregnant women, (b) indirect sex workers, and (c) direct sex workers. The gray curves show the trends of the relative MAE reduction with respect to the original EPP model ($K=0$) as $K$ increases up to 10,000 for each area under each training-test split; the black solid curve shows the median; the black dashed curves show the 2.5th quantile and 97.5th quantile; the red dashed line indicates 0 reduction. In all three sub-populations, not only are the median MAE reductions positive, but also the MAE reductions of most individual test datasets are positive for $25 \le K \le 1000$. As $K$ further increases, there are several test datasets whose MAE increase dramatically, as indicated by negative reductions in the figures. Therefore, we suggest $K$ between 25 and 1,000 for Thailand datasets. Taking $K=500$ as an example, we find the median and (2.5th quantile -- 97.5th quantile) of MAE reductions are 8.4\% and (3.2\%, 17.7\%) in Figure \ref{fig:MAE_Thai} (a), 6.3\% and (1.8\%, 23.9\%) in Figure \ref{fig:MAE_Thai} (b), and 13.6\% and (-3.8\%, 57.0\%) in Figure \ref{fig:MAE_Thai} (c).

Figure 3 shows the estimated prevalence trends among indirect sex workers in four regions of Thailand. The black curves show the posterior median and 95\% credible interval of prevalence trends estimated from EPP without using auxiliary data; the blue curves show the posterior median and 95\% credible interval of prevalence trends estimated from EPP with auxiliary data of sample size $K=500$; the brown curve shows the average prevalence in training data; the red curve shows the average prevalence in test data; the green dots show the auxiliary data. The estimated prevalence trends in North Thailand are similar between EPP models with (blue) and without (black) using auxiliary data. In Central and Northeast Thailand, the auxiliary data (green) reduces the uncertainty of the original EPP estimates (black), and brings the new estimates (blue) closer to the test data average (red), especially for the years that lack of training data (brown). In South Thailand, the original EPP estimates (black) are mostly driven by the prior distribution of parameters in the dynamic system, which implies a gradually increasing and then decreasing prevalence trend with very large uncertainties. The auxiliary data (green) successfully lead to more informative posterior distributions of prevalence trends, and the new estimates (blue) well approximate the test data average (red).

\begin{figure}[!h]
\includegraphics[width=16cm]{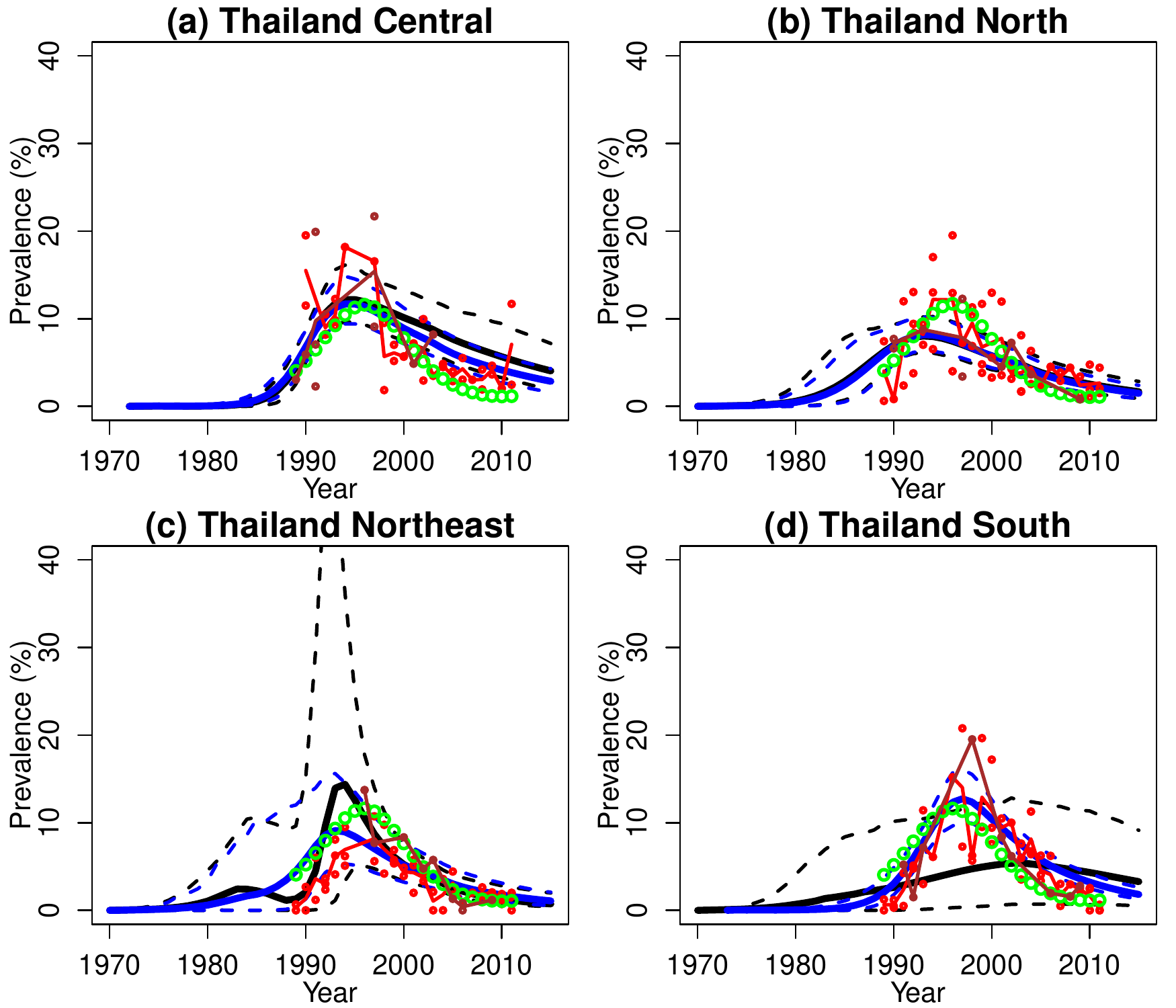}
\caption{\footnotesize{Thailand indirect sex worker example. The black curves show the posterior median and 95\% credible interval of prevalence trends estimated from EPP without using auxiliary data; the blue curves show the posterior median and 95\% credible interval of prevalence trends estimated from EPP by using auxiliary data of sample size $K=500$; the brown curve shows the average prevalence in training data; the red curve shows the average prevalence in test data; the green dots show the auxiliary data.}}
\label{fig:Prev_Thai}
\end{figure}

\section{Discussion}
\label{sect-Discussion}
In this article, we describe an innovative approach that allows the sharing of data information across complicated dynamic systems. The proposed method strengthens the estimation of HIV epidemics at sub-national and sub-population levels, which is critical for local program planning, decision-making, and resource allocation. We demonstrated that our simple pragmatic approach of generating auxiliary data from nearby regions was able to improve both the accuracy and precision of sub-national and sub-group HIV estimates without increasing the computational burden. This method will enable reliable and routine model-based estimates at more granular levels, information urgently needed for guiding and evaluating HIV policy.

The general approach proposed here lends itself to further extensions, where they can be supported by the quality of the data or application to other problems. In this application, we have optimized the weight given to auxiliary data by selecting the parameter $K$ through cross validation. Conceptually, the amount of weight given to the auxiliary data should be related to the amount of similarity in prevalence trends across regions from which information is shared, and it may be that some (more similar) regions should be given greater weight than other regions. This could be incorporated into the GLMM estimation process, for example by specifying a conditional auto-regressive (CAR) error structure for the area-level effects. The approach could also be applicable to incorporating other types of data into dynamic model estimation at lower subnational levels, for example, adult mortality or fertility data into demographic projection models \cite{Wheldon2013}.

We have chosen the natural spline to model the flexible time trend. Other non-parametric models or time series methods are also possible. Alternative model structures in the GLMM could also be explored. In addition, one could consider including reliably-estimated predictors, such as population density, average income, and proportion of migrants. We could also consider spatial dependence in the residuals. Moreover, multiple countries with similar epidemic trends could also be pooled together. In this article, we aim to use a set of relatively simple models so that the audience could focus on the key ideas: separating the data sharing part from the dynamic systems, and incorporating the hierarchical structure through the auxiliary data.


The UNAIDS Reference Group on Estimates, Modelling and Projections has recommended the implementation of the method described in this paper into EPP/Spectrum as an important new feature of the software. In addition, the method is generic enough to model the time trends of other indicators, such as incidence and mortality, upon the data availability. It can also be extended to a wide range of applications, such as cell movements, weather forecasting, demography, etc. 



\section*{Acknowledgments}
This research was supported by the Joint United Nations Programme on HIV/AIDS, NSF grant BCS0941553, and NSF IGERT, DGE1144860. The authors are grateful to Mary Mahy, Keith Sabin, Wiwat Peerapatanapokin and Timothy Hallett for helpful discussions and for sharing data.

\clearpage
\appendix
\makeatletter   
 \renewcommand{\@seccntformat}[1]{APPENDIX~{\csname the#1\endcsname}.\hspace*{1em}}
 \makeatother
 
\section{Model Evaluations for GLMM}
As described in section 3.1, data information on HIV prevalence from different areas and high-risk groups can be shared in the generalized linear mixed model (GLMM), where the time trend is modeled by natural cubic splines. In this appendix, we compare GLMM with alternative models, and evaluate the model performance by using (1) the deviance information criterion (DIC), which is one of the most common metrics for evaluating the performance of Bayesian hierarchical models \cite{Spiegelhalter2002,Spiegelhalter2014}; (2) the mean absolute error (MAE) of the test dataset prediction, where we use the same training/test splits as in Sections 4.1 and 4.2. We use Nigeria as an example of single-risk-group epidemic and Thailand as an example of multiple-risk-group epidemics. 


\subsection{Nigeria Results}
In Nigeria, we model the HIV prevalence among pregnant women in 37 areas. The simplest GLMM under consideration includes the common intercept and spline coefficients for all areas as fixed effects, and site-specific intercepts as random effects, but not any area-specific term. This model allows for site-level correlation in prevalence, but does not allow additional correlation among sites in the same geographic area. We term this the \lq{}national model\rq{}. For another extreme, we consider the \lq{}area model\rq{} that can be approximated by independently fitting the prevalence trend within each area. In the GLMM setting, it includes area-specific intercepts and area-specific spline coefficients, both of which are treated as fixed effects. For the third model, the hierarchical model recommended in Section 3.1 adds area-specific intercepts and area-specific spline coefficients to the \lq{}national model\rq{}, and treats them as random effects; it is a compromise between the \lq{}national model\rq{} and the \lq{}area model\rq{}.

Finally, we explore different settings of the prior distributions assigned to the spline coefficients, $\beta\rq{}s$. The default MCMCglmm assumes a normal distribution with mean zero and standard deviation $10^5$. We stick to the normal priors but explore more informative priors with lower standard deviations.



Appendix Table 1 shows the DIC for Nigeria full data under different models and different prior standard deviations. Under the default prior distribution, the hierarchical model offers the lowest DIC, followed by the area model, and then the national model. For both the national model and the hierarchical model, DIC does not vary much with different standard deviations. The DIC of the area model decreases when smaller prior standard deviations are used. It is because the area model contains too many free parameters, and easily overfits the data; pulling all parameter estimates towards zero prevents overfitting.

\begin{table}[!h]
\caption{The deviance information criterion (DIC) for Nigeria full data (smaller DIC is better). The table compares DIC of 3 models and 7 prior standard deviations.}
\begin{center}
\label{tab:Nigeria_DIC}
\begin{tabular}{|l|c|c|c|c|c|c|c|}
\hline
Prior Standard Deviations & 10 & 20 & 30 & 40 & 50 & 100 & Default \\
\hline
1. National Model & 74835 & 74836 & 74836 & 74836 & 74836 & 74836 & 74836 \\ 
\hline
2. Independent Area Model & 74820 & 74823 & 74824 & 74825 & 74826 & 74828 & 74828 \\ 
\hline
3. Hierarchical Model & 74823 & 74823 & 74824 & 74823 & 74823 & 74823 & 74823 \\ 
\hline
\end{tabular}
\end{center}
\end{table}

Appendix Table 2 shows the MAE for Nigeria test datasets under different models and different prior standard deviations. The hierarchical model offers the lowest MAE regardless the prior standard deviation. Similar to DIC, MAE does not change much with different standard deviations for the national model and the hierarchical model, but decreases for the area model when smaller prior standard deviations are used.

\begin{table}[!h]
\caption{The mean absolute error (MAE) for Nigeria test datasets  (smaller MAE is better). 10 training-test splits are used, and we show the MAE averaged over 10 test datasets. The table compares the MAE of 3 models and 7 prior standard deviations.}
\begin{center}
\label{tab:Nigeria_MAE}
\begin{tabular}{|l|c|c|c|c|c|c|c|}
\hline
Prior Standard Deviations & 10 & 20 & 30 & 40 & 50 & 100 & Default \\
\hline
1. National Model  & .0203 & .0203 & .0203 & .0203 & .0203 & .0203 & .0203 \\ 
\hline
2. Independent Area Model  & .0222 & .0231 & .0241 & .0252 & .0263 & .0296 & .0315 \\ 
\hline
3. Hierarchical Model & .0199 & .0198 & .0199 & .0198 & .0199 & .0198 & .0199 \\ 
\hline
\end{tabular}
\end{center}
\end{table}

According to the full data DIC and the test data MAE, under the default prior, the hierarchical model either performs the best or nearly the best among all candidate models. Therefore, it is recommended for the Nigeria data.

\subsection{Thailand Results}
Thailand has 4 geographic regions: North, Northeast, Central, and South. We also consider 3 different risk groups with reliable data: pregnant women, direct sex workers, and indirect sex workers. On the spatial dimension, we consider the same three options as we did for Nigeria: national, area, and hierarchical. In addition, we also consider three similar options for risk groups: (1) the \lq{}population model\rq{} that treats all risk groups as the same population; (2) the \lq{}group model\rq{} that approximately fits each risk group separately; (3) the hierarchical model that introduces the group-specific parameters as random effects. All together, we compare 9 different models. 

Similarly to the Nigeria example, we recommended the hierarchical structure for the spatial dimension. However, it is known that the epidemic patterns are often quite different among sub-populations with various risk behaviors. We recommended not assuming any similarity of the group-specific parameters. 
In Appendix Table 3, for the 9 models, we compare the full data DIC, average test data MAE, and standard deviation of test data MAE. The default prior distribution was used. The recommended model with hierarchical spatial structure and fixed group effects (Model 8) either performs the best or nearly the best among all candidate models. 

\begin{table}[!h]
\caption{The deviance information criterion (DIC) for Thailand full data, and the mean absolute error (MAE) for Thailand test data. For the test data MAE, 10 random training-test splits were considered, and we show the mean and standard deviation across 10 random splits.}
\begin{center}
\label{tab:Thailand}
\begin{tabular}{|lll|c|c|}
\hline
Model ID & Spatial & Sub-Population & Full Data DIC & Test Data MAE: Mean (Std) \\
\hline
1 & national & population & 394092 & .0579 (.0026) \\ 
\hline
2 & national & group & 393399 & .0302 (.0030) \\ 
\hline 
3 & national & hierarchical  & 393422 & .0314 (.0030) \\ 
\hline
4 & area & population & 394085 & .0613 (.0034) \\ 
\hline
5 & area & group & 393356 & .0340 (.0032) \\ 
\hline
6 & area & hierarchical & 393374 & .0355 (.0027) \\
\hline 
7 & hierarchical & population & 394086 & .0579 (.0026) \\ 
\hline
8 & hierarchical & group & 393358 & .0302 (.0031) \\ 
\hline
9 & hierarchical & hierarchical  & 393376 & .0329 (.0032) \\ 
\hline
\end{tabular}
\end{center}
\end{table}

\clearpage

\bibliographystyle{ECA_jasa}
\bibliography{EPP}

\end{document}